\documentclass{aa}
\usepackage{subfig}
\usepackage{amsmath}
\usepackage{times}
\usepackage{natbib}
\usepackage[colorlinks,linkcolor=blue,anchorcolor=blue,citecolor=blue]{hyperref}
\usepackage[hyperref]{backref}

\usepackage{graphicx}

\usepackage{color}
\usepackage{multirow}

\begin{document}

\title{Measuring redshifts using X-ray spectroscopy of galaxy
  clusters: results from Chandra data and future prospects}
\titlerunning{Measuring redshifts using X-ray spectroscopy of galaxy
  clusters}

\author{H. Yu\inst{1,2}, P. Tozzi\inst{2,3}, S. Borgani\inst{4,3},
  P. Rosati\inst{5}, Z.-H. Zhu\inst{1}}

\authorrunning{H.Yu et al.}

\offprints{H. Yu \email{heng@oats.inaf.it}}

\institute{Department of Astronomy, Beijing Normal University, Beijing
  100875, China \and INAF Osservatorio Astronomico di Trieste, via
  G.B. Tiepolo 11, I--34143, Trieste, Italy \and INFN-- National
 Institute for Nuclear Physics, Trieste, Italy \and Dipartimento di
  Fisica dell'Universit\`a di Trieste, via G.B. Tiepolo 11, I--34131,
  Trieste, Italy \and ESO - European Southern Observatory,
Karl-Schwarzschild Str. 2, D-85748 Garching b. Munchen, Germany}


\abstract 
{The ubiquitous presence of the Fe line complex in the X-ray spectra
  of galaxy clusters offers the possibility of measuring their
  redshift without resorting to spectroscopic follow-up observations.  In
  practice, the blind search of the Fe line in X-ray spectra is a
  difficult task and is affected not only by limited
  S/N (particularly at high redshift), but also by
  several systematic errors, associated with varying Fe abundance values,
  ICM temperature gradients, and instrumental characteristics.}
{ We assess the accuracy with which the redshift of galaxy
  clusters can be recovered from an X-ray spectral analysis of Chandra
  archival data.  We present a strategy to compile large surveys 
  of clusters whose identification and redshift measurement are both 
  based on X-ray data alone.  }
{We apply a blind search for K--shell and L--shell Fe line complexes in
  X-ray cluster spectra using Chandra archival observations of galaxy
  clusters.  The Fe line can be detected in the ICM 
  spectra  by simply analyzing the C-statistics variation $\Delta
  C_{stat}$ as a function of the redshift parameter, when all the
  other model parameters are frozen to the best--fit values.
  We repeat the measurement under different conditions, and compare the
  X-ray derived redshift $z_X$ with the one obtained by means of optical
  spectroscopy $z_o$.  We explore how a number of priors 
  on metallicity and luminosity can be effectively used to reduce
  catastrophic errors. The $\Delta C_{stat}$ provides the most
  effective means for discarding wrong redshift measurements and 
  estimating the actual error in $z_X$.  }
{We identify a simple and efficient procedure for optimally measuring the
  redshifts from the X-ray spectral analysis of clusters of galaxies.
  When this procedure is applied to mock catalogs extracted from high
  sensitivity, wide-area cluster surveys, such as those proposed with
  Wide Field X-ray Telescope (WFXT) mission, it is possible to obtain
  complete samples of X-ray clusters with reliable redshift
  measurements, thus avoiding time-consuming optical spectroscopic
  observations.  Our analysis shows that, in the case of WFXT, a blind
  Fe line search is 95\% successful for spectra with more than 1000
  net counts, whenever $\Delta C_{stat} > 9$, corresponding
  formally to a 3$\sigma$ confidence level.  The average error in the
  redshift $z_X$ decreases rapidly for higher values of $\Delta
  C_{stat}$. Finally, we discuss how to estimate the completeness of
  a large cluster samples with measured $z_X$.  This methodology will
  make it possible to trace cosmic growth by studying the evolution of the
  cluster mass function directly using X-ray data.}
{}


\maketitle

\section{Introduction}

The study of clusters of galaxies allows one to investigate the 
large-scale structure of the Universe, constrain the cosmological
parameters and the spectrum of the primordial density fluctuations,
and study the interactions between the member galaxies and the
ambient intra cluster medium \citep[ICM, see
][]{2002Rosati,2005Schuecker,2005Voit,2006Borgani,2007Tozzi,
  2009Vikhlinin,2010Mantz,2010Bohringer}.  The X--ray band is an
optimal observational window for identifying and studying clusters of
galaxies.  The thermal bremsstrahlung emission due to the hot diffuse
ICM is roughly proportional to the baryon density squared. This makes
clusters of galaxies prominent sources in the X-ray sky.

Thanks to the spatial resolution of Chandra and the
high sensitivity of XMM, coupled with their excellent spectral
resolutions, it has become possible to study the detailed
thermodynamical properties of the ICM.  In particular, the detection
of emission lines of highly ionized metals, in both the core and 
outer regions of groups and clusters, has proven to be very efficiency 
in investigating the chemical properties of the ICM.  Indeed, the
ubiquitous presence of the K--shell Fe line complex at 6.7-6.9 keV,
which was first detected by \citet{1976Mitchell}, has been
detected at the highest redshift $z\sim 1.4$ where X-ray
clusters have been observed \citep[see][]{2004Rosati,stanf05,2009Rosati}.
In principle, abundance measurements based on the detection of the
Fe line also provide a way to measure the position of the Fe line, 
thus the cluster redshift.

A few redshifts have been measured using X-ray spectral analyses
in cases of no previous optical spectroscopic observations
\citep{2004Hashimoto,2008Lamer}.  But, this approach has never been
used systematically in cluster surveys.\footnote{We note that a
blind survey of the 6.4 Fe line has been applied to AGN X-ray
surveys with interesting results \citep{2004Maccacaro,2005Braito}.
The search of line emission in deep AGN surveys, however, is
significantly different with respect to the search of Fe line
emission in the thermal spectra of ICM, and their method is
unsuitable in our case.}  The main reason is that most of the
existing cluster surveys are based on source samples selected by
ROSAT, whose energy range is 0.1-2.4 keV, hence does not
cover the hard band where the Fe lines lie. To date, most ROSAT
clusters have been confirmed through optical imaging and spectroscopic
observations.  Several studies based on Chandra and XMM data
\citep{2003Tozzi,2004Ettori,2007Balestra,2008Maughan} simply confirmed
the excellent agreement of X-ray and optical redshifts.  This
agreement was achieved using the optical redshift as the initial 
redshift parameter in the X-ray spectral fit.  The blind search of the redshift
from the X-ray spectral analysis has never been explored thoroughly.

The situation could change significantly with the next-generation
X--ray surveys of planned future missions sensitive in the 0.5-10 keV
band, such as eROSITA \citep{2010Predehl} or the proposed Wide Field
X-ray Telescope \citep[WFXT, see ][]{2010WFXT}. In these surveys, we
expect to detect several hundred thousands of groups and clusters of
galaxies, making an {\sl ad hoc} follow-up program infeasible, and
requiring careful coordination with existing and future optical and
IR surveys \citep{2010WFXT2}.  Therefore, the ability to recover
redshifts on the basis of the X-ray data alone, would be critical since
it would allow one to measure the redshift, thus the intrinsic
physical properties of ICM, such as the X--ray luminosity $L_X$, 
ICM temperature $T_X$, iron abundance of the ICM $Z$, and the
electron density $n_e$. In this case, it will be not only possible
to investigate the physics of the ICM, but also perform
cosmological tests with large cluster surveys entirely characterized
on the basis of X-ray data.

The factors which affect the measurement of redshift from X-ray
spectra are: the signal-to-noise ratio, hereafter S/N (or the total number of
detected counts), the energy dependence of the instrument effective
area, its spectral resolution, the intrinsic Fe abundance (or
metallicity in general), the ICM temperature, and the 
actual redshift.  A possible departure from collisional equilibrium, or
the incompleteness of the atomic models used in the fitting procedure,
are not investigated here, and are expected to be negligible in the
analysis of high-z cluster spectra.  In this work, we use the data
archive of the Chandra X-ray satellite to investigate the ability of
recovering the X-ray redshift $z_X$ with a blind search of X-ray
emission lines. This can lead to interesting applications for future X-ray
survey missions.

The paper is organized as follows: in Section 2, we describe the data we use
for our study.  In Section 3, we present a simple and efficient strategy to
search for emission lines in X-ray spectra.  In Section 4, we explore the
use of weak priors to test their effect on X-ray redshift measurements.
In Section 5, we apply our algorithm to the expected results from the WFXT
surveys.  Finally in Section 6, we summarize our conclusions.  Throughout
the paper, we assume the 7-year WMAP cosmology $\Omega_{\Lambda} = 0.728$,
 $\Omega_m =0.272$, and $H_0 = 70.4 $ km s$^{-1}$ Mpc$^{-1}$ \citep{Kom2010}.

\section{The data}

We searched the Chandra archive for targeted observations of galaxy
clusters with different temperatures and redshifts.  We adopted a
sample of 46 clusters (see Table \ref{46}) mostly based on those studied
by \citet{2007Balestra}.This sample includes clusters with
temperatures from $\sim 4$ keV to $\sim 14$ keV over a wide redshift
range ($0.16 < z < 1.4$).  The sample is incomplete and represents a
collection of targets from different surveys (see reference list in
Table 1).   We also remark that in several cases we did not make use
of all the available archival observations, in order to explore a
wider S/N range. However, the most important point is
that this sample covers a sizable portion of 
redshift-temperature space allowing us to comprehensibly explore the ability to
measure the redshifts from the X-ray spectra.

\begin{figure}
\includegraphics[width=0.5\textwidth]{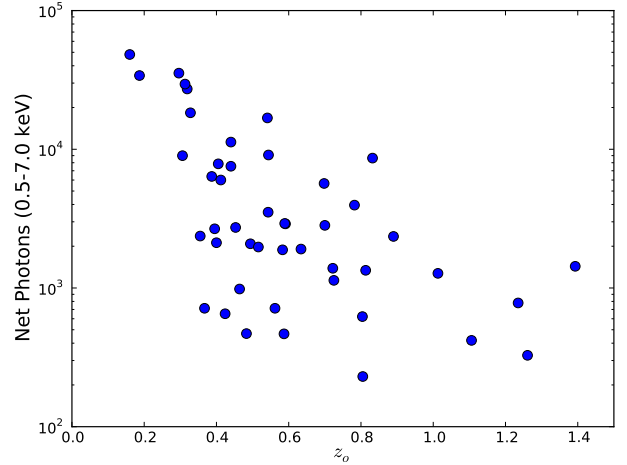} 
\caption{\label{ots}Detected photons as a function of redshift for the
  46 clusters selected from the Chandra archive.  The wide range of
  values reflects the different fluxes of each object and the different
  exposure times of the available data.}
\end{figure}

The number of detected photons in the 0.5-7 keV band ranges from $\sim
230$ to $\sim 45000$ as shown in Figure \ref{ots}.  Most were
gathered with Chandra ACIS-I in the FAINT/VERYFAINT mode, while only six
clusters were observed with ACIS-S.  Calibration files were obtained
using the most recent release of CALDB at the time of writing (CALDB
4.2). Image reduction began from the level 1 event file. We applied a
charge transfer inefficiency (CTI) correction, FAINT/VERYFAINT
cleaning, grade correction, and time-dependent gain correction. High
background intervals were filtered with a $3\sigma$ clipping
procedure. The response matrix files (RMF) and the ancillary response
files (ARF), necessary for the X-ray spectral analysis, were generated
with {\tt mkacisrmf} and {\tt mkwarf}. For details about data reduction,
we refer to \citet{2007Balestra}.

The spectra were analyzed with XSPEC v12.6.0 \citep{arn96}.  Data were
fitted with a single-temperature {\tt mekal} model
\citep{85Mewe,86Mewe,kaa92,lie95} in which the ratio of the
elements was fixed to the solar value as in \citet{1989Ge}. Galactic
absorption is modeled with {\tt tbabs} \citep{wil00}.  The optical
redshifts were collected from the literature. The typical error in the
optical redshift is $\Delta z \sim 0.003$.  We adopt the optical
redshift when measuring the temperatures and metallicities of the ICM.
Our results, shown in Table \ref{46}, differ slightly from
previous analyses presented in \citet{2007Balestra} because of the updated
Chandra calibration and in some cases the different extraction
region.  Computed errors always correspond to the $1\sigma$ confidence
level.

\begin{table*}
\centering\begin{tabular}{lllcclllll}
\hline
ID & $z_{o}$ & ref & ObsID & Detector/Mode & $t_{exp}(ks)$ & $T(z_o)(keV)$ & $Z_{Fe}(z_o)/Z_{\odot}$ & $nH(cm^{-2})$ \\ 
\hline 
Abell907 & 0.160 & 1 & 3185 & I-V & 47 & $5.56_{-0.09}^{+0.09}$ & $0.49_{-0.04}^{+0.04}$ & 5.65$\times 10^{20}$\\ 
Abell1689 & 0.187 & 2 & 540,1663 & I-F & 21 & $13.74_{-0.65}^{+0.65}$ & $0.34_{-0.07}^{+0.07}$ & 1.82$\times 10^{20}$\\ 
1E0657-56 & 0.296 & 3 & 554 & I-F & 25 & $13.17_{-0.63}^{+0.63}$ & $0.2_{-0.06}^{+0.06}$ & 6.5$\times 10^{20}$\\ 
MS1008.1-1224 & 0.306 & 4 & 926 & I-V & 44 & $6.09_{-0.33}^{+0.32}$ & $0.27_{-0.06}^{+0.06}$ & 7.26$\times 10^{20}$\\ 
MS2137.3-2353 & 0.313 & 5 & 928 & S-V & 34 & $5.26_{-0.11}^{+0.12}$ & $0.35_{-0.03}^{+0.04}$ & 3.55$\times 10^{20}$\\ 
Abell1995 & 0.319 & 6 & 906 & S-F & 56.4 & $9.13_{-0.31}^{+0.32}$ & $0.4_{-0.06}^{+0.07}$ & 1.42$\times 10^{20}$\\ 
ZwCl1358.1+6245 & 0.328 & 8 & 516 & S-F & 48.3 & $7.12_{-0.25}^{+0.26}$ & $0.39_{-0.07}^{+0.07}$ & 1.92$\times 10^{20}$\\ 
MACSJ0404.6+1109 & 0.355 & 9 & 3269 & I-V & 21.6 & $6.9_{-0.8}^{+0.6}$ & $0.16_{-0.11}^{+0.07}$ & 1.43$\times 10^{20}$\\ 
RXJ0027.6+2616 & 0.367 & 9 & 3249 & I-V & 9.8 & $8.07_{-1.39}^{+1.73}$ & $0.52_{-0.24}^{+0.30}$ & 3.86$\times 10^{20}$\\ 
MACSJ1720.2+3536 & 0.387 & 10 & 3280 & I-V & 20.8 & $6.54_{-0.34}^{+0.35}$ & $0.45_{-0.08}^{+0.08}$ & 3.4$\times 10^{20}$\\ 
ZwCl0024.0+1652 & 0.395 & 11 & 929 & S-F & 39.5 & $4.57_{-0.27}^{+0.49}$ & $0.75_{-0.18}^{+0.21}$ & 4.23$\times 10^{20}$\\ 
RXJ1416.4+4446 & 0.400 & 12 & 541 & I-V & 31 & $3.56_{-0.19}^{+0.2}$ & $0.86_{-0.21}^{+0.25}$ & 1.22$\times 10^{20}$\\ 
MACSJ0159.8-0849 & 0.405 & 13 & 3265 & I-V & 17.6 & $9.53_{-0.5}^{+0.74}$ & $0.35_{-0.08}^{+0.08}$ & 2.08$\times 10^{20}$\\ 
MACSJ2228.5+2036 & 0.412 & 9 & 3285 & I-V & 20 & $8.25_{-0.6}^{+0.61}$ & $0.35_{-0.09}^{+0.09}$ & 4.58$\times 10^{20}$\\ 
MS0302.7+1658 & 0.424 & 8 & 525 & I-V & 10 & $4.38_{-0.45}^{+0.61}$ & $0.45_{-0.19}^{+0.24}$ & 10.9$\times 10^{20}$\\ 
MACSJ0417.5-1154 & 0.44 & 14 & 3270 & I-V & 12 & $12.48_{-1.06}^{+1.24}$ & $0.29_{-0.11}^{+0.11}$ & 3.86$\times 10^{20}$\\ 
MACSJ1206.2-0847 & 0.44 & 15 & 3277 & I-V & 23 & $12.61_{-0.88}^{+0.97}$ & $0.17_{-0.08}^{+0.08}$ & 3.72$\times 10^{20}$\\ 
RXJ1701.3+6414 & 0.453 & 16 & 547 & I-V & 49 & $4.49_{-0.26}^{+0.3}$ & $0.51_{-0.06}^{+0.13}$ & 2.46$\times 10^{20}$\\ 
RXJ1641.8+4001 & 0.464 & 16 & 3575 & I-V & 45 & $4.81_{-0.54}^{+0.62}$ & $0.48_{-0.17}^{+0.21}$ & 1.1$\times 10^{20}$\\ 
MACSJ1824.3+4309 & 0.483 & 10 & 3255 & I-V & 14.8 & $8.94_{-2.26}^{+3.57}$ & $<0.23$ & 4.58$\times 10^{20}$\\ 
MACSJ1311.0-0311 & 0.494 & 17 & 3258 & I-V & 14.8 & $8.67_{-0.91}^{+0.98}$ & $0.37_{-0.14}^{+0.15}$ & 1.87$\times 10^{20}$\\ 
RXJ1524.6+0957 & 0.516 & 16 & 1664 & I-V & 50 & $5.58_{-0.48}^{+0.59}$ & $0.34_{-0.12}^{+0.14}$ & 2.91$\times 10^{20}$\\ 
MS0015.9+1609 & 0.541 & 8 & 520 & I-V & 67 & $8.3_{-0.32}^{+0.32}$ & $0.31_{-0.05}^{+0.05}$ & 7.26$\times 10^{20}$\\ 
MACSJ1423.8+2404 & 0.543 & 18 & 1657 & I-V & 18.5 & $7.61_{-0.52}^{+0.68}$ & $0.3_{-0.09}^{+0.1}$ & 2.38$\times 10^{20}$\\ 
MACSJ1149.5+2223 & 0.544 & 18 & 1656, 3589 & I-V & 38 & $12.75_{-0.97}^{+1.17}$ & $0.27_{-0.1}^{+0.1}$ & 2.28$\times 10^{20}$\\ 
SC1120-1202 & 0.562 & 7 & 3235 & I-V & 68 & $6.49_{-1.11}^{+1.35}$ & $0.21_{-0.19}^{+0.23}$ & 5.19$\times 10^{20}$\\ 
MS2053.7-0449 & 0.583 & 8 & 551, 1667 & I-V & 88 & $6.94_{-0.62}^{+0.67}$ & $0.18_{-0.1}^{+0.12}$ & 4.96$\times 10^{20}$\\ 
RXJ0956.0+4107 & 0.587 & 16 & 5294 & I-V & 17.2 & $8.19_{-1.79}^{+2.78}$ & $0.18_{-0.18}^{+0.32}$ & 1.14$\times 10^{20}$\\ 
MACSJ2129.4-0741 & 0.589 & 18 & 3199 & I-V & 17.6 & $9.48_{-0.8}^{+1.25}$ & $0.54_{-0.14}^{+0.15}$ & 4.86$\times 10^{20}$\\ 
MACSJ0647.7+7015 & 0.591 & 18 & 3196 & I-V & 19.2 & $12.10_{-1.08}^{+1.54}$ & $<$0.1 & 5.64$\times 10^{20}$\\ 
RXJ0542.8-4100 & 0.634 & 7 & 914 & I-F & 50 & $7.93_{-0.85}^{+1.11}$ & $0.18_{-0.12}^{+0.14}$ & 3.73$\times 10^{20}$\\ 
MACSJ0744.9+3927 & 0.698 & 18 & 3197, 3585 & I-V & 40 & $9.51_{-0.54}^{+0.77}$ & $0.29_{-0.08}^{+0.09}$ & 5.7$\times 10^{20}$\\ 
RXJ1221.4+4918 & 0.70 & 12 & 1662 & I-V & 78 & $9.2_{-0.83}^{+1.06}$ & $0.29_{-0.12}^{+0.14}$ & 1.47$\times 10^{20}$\\ 
RXJ2302.8+0844 & 0.722 & 19 & 918 & I-F & 108 & $7.35_{-0.85}^{+1.22}$ & $0.17_{-0.15}^{+0.17}$ & 4.91$\times 10^{20}$\\ 
RXJ1113.1-2615 & 0.725 & 19 & 915 & I-F & 103 & $6.41_{-0.82}^{+0.89}$ & $0.35_{-0.17}^{+0.19}$ & 5.48$\times 10^{20}$\\ 
MS1137.5+6624 & 0.782 & 20 & 536 & I-V & 117 & $7.33_{-0.51}^{+0.58}$ & $0.24_{-0.11}^{+0.12}$ & 1.2$\times 10^{20}$\\ 
RXJ1350.0+6007 & 0.804 & 21 & 2229 & I-V & 58 & $4.38_{-0.57}^{+0.76}$ & $0.57_{-0.26}^{+0.34}$ & 1.78$\times 10^{20}$\\ 
RXJ1317.4+2911 & 0.805 & 21 & 2228 & I-V & 110.5 & $3.77_{-0.71}^{+1.09}$ & $0.48_{-0.36}^{+0.55}$ & 1.1$\times 10^{20}$\\ 
RXJ1716.4+6708 & 0.813 & 22 & 548 & I-F & 51 & $6.57_{-0.67}^{+0.72}$ & $0.57_{-0.17}^{+0.14}$ & 3.7$\times 10^{20}$\\ 
MS1054.4-0321 & 0.832 & 23 & 512 & S-F & 80 & $7.26_{-0.35}^{+0.44}$ & $0.23_{-0.07}^{+0.07}$ & 3.6$\times 10^{20}$\\ 
1WGAJ1226.9+3332 & 0.89 & 24 & 3180 & I-V & 31.5 & $12.03_{-1.17}^{+1.46}$ & $<$0.11 & 1.38$\times 10^{20}$\\ 
CLJ1415.1+3612 & 1.013 & 19 & 4163 & I-V & 89 & $6.76_{-0.67}^{+0.74}$ & $0.32_{-0.14}^{+0.16}$ & 1.09$\times 10^{20}$\\ 
RXJ0910+5422 & 1.106 & 25 & 2227, 2452 & I-V & 170 & $5.85_{-1.08}^{+1.59}$ & $0.08_{-0.08}^{+0.26}$ & 2.2$\times 10^{20}$\\ 
RXJ1252-2927 & 1.235 & 27 & 4198, 4403 & I-V & 188.4 & $6.81_{-0.87}^{+1.24}$ & $0.84_{-0.27}^{+0.32}$ & 5.95$\times 10^{20}$\\ 
RXJ0848.9+4452 & 1.261 & 26 & 927, 1708 & I-V & 184.5 & $4.44_{-0.86}^{+1.11}$ & $0.37_{-0.29}^{+0.47}$ & 2.6$\times 10^{20}$\\ 
XMMUJ2235.3-2557 & 1.393 & 28 & 6975/6, 7367/8, 7404 & S-V & 196 & $9.96_{-1.28}^{+1.6}$ & $0.41_{-0.21}^{+0.24}$ & 1.47$\times 10^{20}$\\ 
\hline
\end{tabular}

\caption{\label{46}The cluster sample used in this work, in order of increasing redshift.  We show the
  properties of each data-set, including the references to previously
  published works, observation ID, detector (I=ACIS-I, S=ACIS-S) ,
  observation mode (F=FAINT, V=VERYFAINT), and exposure time.  The
  best-fit parameters $T(z_o)$ and $Z_{Fe}(z_o)$ are obtained with the redshift
  frozen to the optical value $z_o$.  Temperature and metal abundance
  refer to the global fit to the spectrum extracted within the radius
  $R_{ext}$, and therefore represents an average value whenever a
  temperature or a metallicity gradient is present.  The last column
  refers to the Galactic Hydrogen column density as measured by
  \citet{2005LAB}.}  \tablebib{ (1):\citet{rsA907}; (2):\citet{rsA1689};
  (3):\citet{rsE0657}; (4):\citet{rsMS1008}; (5):\citet{rsA2137};
  (6):\citet{rsAbell}; (7):\citet{2007Balestra}; (8):\citet{1991EMS};
  (9):\citet{2000ROSAT}; (10):\citet{rsM1720}; (11):\citet{rs0024};
  (12):\citet{rs1416}; (13):\citet{rs0159}; (14):\citet{rs0417};
  (15):\citet{rs1206}; (16):\citet{rs1701}; (17):\citet{2004Allen};
  (18):\citet{rs1149}; (19):\citet{rs1113}; (20):\citet{1994Gioia};
  (21):\citet{2002Holden}; (22):\citet{1997Henry};
  (23):\citet{2007Tran}; (24):\citet{2001Ebeling};
  (25):\citet{2002Stanford}; (26):\citet{1999Rosati};
  (27):\citet{2004Rosati}; (28):\citet{2005Mullis}.}

\end{table*}

\section{Blind search for the redshift with no priors}

\subsection{Method}

The $K_\alpha$ shell complex of Fe consists of two groups
of lines. The first is the He-like iron (Fe XXV) K-shell
complex, whose resonant line is at 6.7 keV, that is the most
prominent spectral feature of ICM spectra at high temperatures
(above 2 keV).  The second group corresponds to the H-like iron 
(Fe XXVI) spectral line whose
resonant line is at 6.9 keV.  Both groups include lines
corresponding to several transitions as predicted in the case of
collisional ionization equilibrium.  The H-like and He-like
complexes, separated by about 0.2 keV, can be resolved at the
spectral resolution of Chandra ACIS in data of high S/N;
in this case, the line ratio can be used as a temperature diagnostic
in addition to the continuum shape.  On the other hand, the
structure of each group is not resolved and reflects the
asymmetric shape of the observed feature.  Both effects are taken
into account by the {\tt mekal} model, which includes excitation,
radiative recombination, dielectronic recombination, and inner shell
ionization, and assumes the optically thin limit (i.e., that no
photo-ionization or photo-excitation effects are taken into
account).  Given the low S/N of distant cluster spectra, the
$K_\alpha$ line complex often appears as a single, prominent
feature. 

Owing to uncertainties in the ACIS calibration below 0.5 keV, we excluded
these low energy photons from the spectral analysis to avoid any
systematic bias.  The effective cut at high energies is generally
around $7$ keV, since the S/N of a thermal spectrum rapidly decreases
above $5$ keV.  Therefore, we consider the energy range 0.5-7 keV for both
imaging and spectral analysis.  Given the explored redshift
interval ($0<z<2$), the Fe line complex is always well within this
range.

Background is usually selected in the same chip where the source lies.
When several observations are used, the background can only be
selected in overlapping regions.  For some nearby clusters that
occupy a whole chip, or even an entire field (Abell 1689 for
example), we compile a synthetic background using CALDB.  The centroid
of the cluster emission is determined by surface brightness fits
adopting a standard $\beta$ model \citep{cav76}. A set of
circular regions are then drawn to select the extraction radius
$R_{ext}$ that maximizes the S/N, computed as
$SN(R) = (S(R)- B)/\sqrt{S(R) + (A_S/A_B)\times B}$, where $S(R)$
are the total number of photons detected within the radius $R$, and B are the
background number of photons expected in the same area. The factor $A_S/A_B$
is the geometrical backscale, i.e. the ratio of the area of the
source extraction region to that of the background.  Since our
sources are extended and the background is chosen to be as
close as possible to the source, this factor can be of order $\leq
1$.  This expression properly takes into account the statistical
uncertainty in the number of photons in the source region and the
region where the background is sampled. The extraction regions are
selected on the X-ray image in the total band (0.5-7 keV).

\begin{figure}[h]
\includegraphics[width=0.5\textwidth]{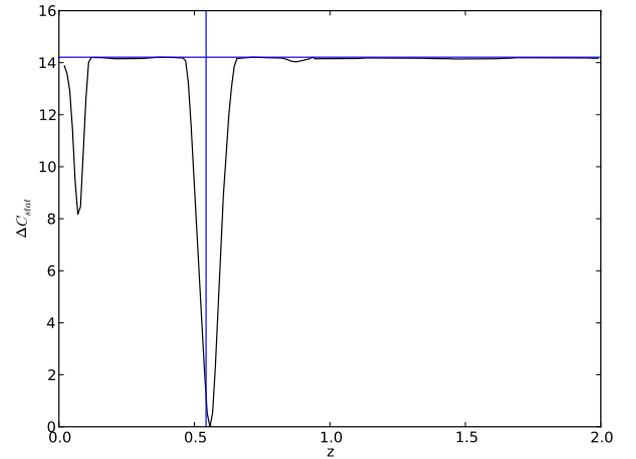}
\caption{$\Delta C_{stat}$ as a function of the redshift parameter for
  MACSJ1423. The vertical line indicates the optical redshift.  The
  horizontal line corresponds to the minimum $C_{stat}$ allowed when
  the metal abundance is set to zero. }
\label{cstat} 
\end{figure}

We adopt the default Levenberg-Marquardt fitting algorithm.  There are
four parameters to be fitted: temperature, metal abundance, redshift,
and normalization.To test the capability of recovering the
actual redshift with a blind search of the K$_\alpha$ Fe line in the
X--ray spectrum, we repeat our fits starting from a reference set of
parameters, corresponding to $z_{start}=0.1$, $kT_{start}=5$ keV, and
$Z_{start}/Z_\odot = 0.3$ in units of \citet{1989Ge} for all the
clusters.  We use Cash statistics \citep{1979Cash} applied to an unbinned
source plus background counts, and therefore exploit the full spectral
resolution of the ACIS-I and ACIS-S instrument.  Cash statistics
ensure better performance with respect to the canonical $\chi ^2$
analysis of binned data, particularly for low S/N spectra
\citep{nou89}.  After finding the absolute minimum, we explore the
redshift space with the {\tt steppar} command, covering the entire
range of possible values from $z=0.01$ to $z=2.0$ with a very small
step $\delta z = 0.01$.  When a new minimum is obtained, the best fit
is automatically updated. We then plot the difference of the C-stat
value with respect to the minimum as a function of redshift.  The
$\Delta C_{stat}(z)$ function is therefore the difference between the
absolute minimum and the best-fit value obtained when the $z$
parameter does not match any line, which is equivalent to optimizing the
fit when the Fe abundance is forced to be zero.  One example
(MACSJ1423) is shown in Figure \ref{cstat}.  The $\Delta C_{stat}$
rapidly declines to a minimum whenever a Fe line candidate is
found.  The deepest minimum in the C-statistics as a function of the
redshift in Figure \ref{cstat} clearly shows that the best-fit
redshift agrees with the optical redshift (indicated by the
vertical line).  The horizontal line, therefore, corresponds to the
minimum $C_{stat}$ value obtained with zero metal abundance.

\begin{figure}[h]
\includegraphics[width=0.5\textwidth]{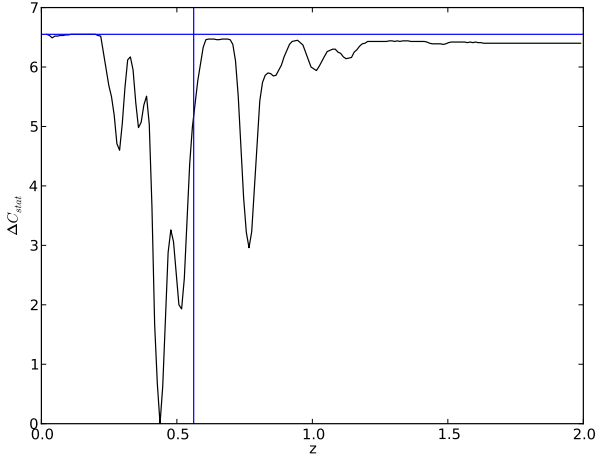}
\caption{$\Delta C_{stat}$ as a function of the redshift parameter for
  SC1120. The vertical line indicates the optical redshift.  At
  variance with Figure \ref{cstat}, the search for a minimum in
  $C_{stat}$ leads to a catastrophic error. }
\label{cstat2} 
\end{figure}

The case of a catastrophic error is shown in Figure \ref{cstat2}.
Here there are no minima corresponding to the optical redshift.  We
note that, instead, there are several comparable secondary minima.
We explore a possible use of secondary minima in Section 3.7.

\subsection{Results}

\begin{figure}
\includegraphics[width=0.5\textwidth]{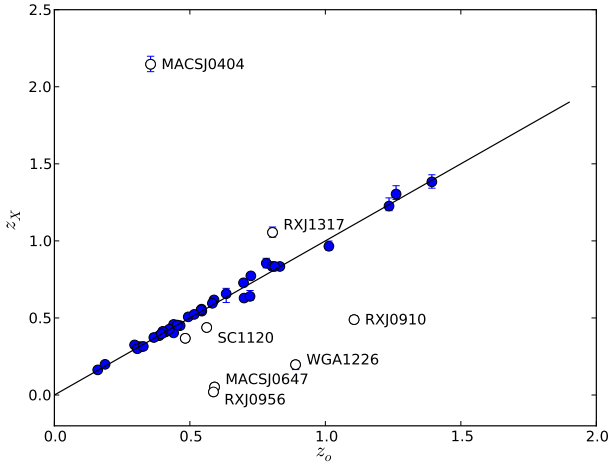}
\caption{\label{OX}X-ray redshift $z_X$ compared to the optical values
  $z_o$ for the 46 Chandra clusters in our sample.  The eight empty
  circles are the catastrophic failures selected with a $3\sigma$
  clipping.  }
\end{figure}

\begin{figure}
\includegraphics[width=0.5\textwidth]{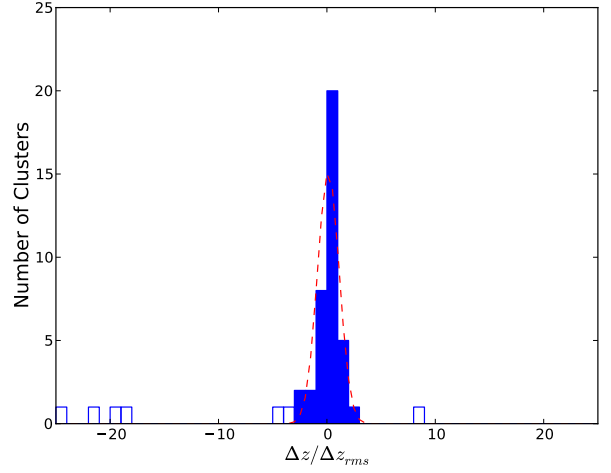} 
\caption{\label{hst}Histogram of the quantity $\Delta z/\Delta
  z_{rms}$.  Catastrophic failures, identified with a $3\sigma$
  clipping, are shown with an empty histogram.  A Gaussian fit to the
  data is shown as a dashed line.}
\end{figure}

For most clusters, the redshifts $z_X$ found from the X-ray
spectral analysis are consistent with their optical values $z_o$
within the errors as shown in Figure \ref{OX}.  However, there are
still several clusters for which $z_X$ is far from $z_o$. To define
these "catastrophic failures", we perform a $3\sigma$-clipping by
computing the rms value $\Delta z_{rms} \equiv
\sqrt{\Sigma(z_X-z_o)^2/n}$ for the entire sample of 46 clusters.  The
catastrophic errors are shown as empty circles in Figure \ref{OX} and
as an empty histogram in Figure \ref{hst}, where we plot the quantity
$\Delta z/\Delta z_{rms}$.  Thanks to this definition, we identify eight
catastrophic errors.  After removing the catastrophic failures, the
rms redshift deviation is $\Delta z \sim 0.03$.  We note that
this uncertainty is somewhat larger than the typical statistical
error $\sigma_z$ estimated by the XSPEC fitting routine.  We argue
that the statistical error in the redshift is slightly
underestimated.  As we see later (Section 5.1), $\Delta C_{stat}$
is not only the most relevant parameter to accept or reject
the $z_X$ value, but can also be used to adjust the statistical
error obtained by the spectral fit thanks to its strong correlation
with the actual value of $z_X-z_o$.

Our results for the entire sample of {\sl Chandra} clusters is shown in
Table \ref{fit}.  From a comparison with Table \ref{46}, it is
possible to see that, apart from the catastrophic failures, the
best-fit temperatures are unaffected, while the best-fit metal
abundances are systematically higher (see discussion in Section 3.5).  

\begin{table*}
\centering\begin{tabular}{l|llllcrrl}
\hline
ID & $z_{o}$ & $z_X$ & $T(z_X)$ (keV) & $Z_{Fe}(z_X)/Z_{\odot}$ & Cts (0.5-7 keV) & SN & $R_{ext}('')$ & $\Delta C_{stat}$  \\ 
\hline 
Abell907 & 0.160 & $0.163_{-0.002}^{+0.002}$ & $5.58_{-0.1}^{+0.09}$ & $0.49_{-0.04}^{+0.04}$ & 48230$\pm$238 & 202.5 & 162.4 & 220.56\\ 
Abell1689 & 0.187 & $0.199_{-0.005}^{+0.005}$ & $13.90_{-0.65}^{+0.64}$ & $0.37_{-0.07}^{+0.07}$ & 34008$\pm$200 & 170.3 & 93.5 & 31.59\\ 
1E0657-56 & 0.296 & $0.325_{-0.006}^{+0.006}$ & $13.3_{-0.63}^{+0.63}$ & $0.26_{-0.05}^{+0.05}$ & 35384$\pm$204 & 173.1 & 211.6 & 24.65\\ 
MS1008.1-1244 & 0.306 & $0.3_{-0.016}^{+0.014}$ & $6.05_{-0.34}^{+0.37}$ & $0.27_{-0.06}^{+0.06}$ & 8993$\pm$103 & 87.2 & 108.2 & 24.32\\ 
MS2137.3-2353 & 0.313 & $0.313_{-0.002}^{+0.002}$ & $5.26_{-0.12}^{+0.11}$ & $0.35_{-0.03}^{+0.04}$ & 29509$\pm$176 & 167.9 & 76.3 & 171.03\\ 
Abell1995 & 0.319 & $0.313_{-0.005}^{+0.003}$ & $9.05_{-0.31}^{+0.32}$ & $0.41_{-0.06}^{+0.07}$ & 27217$\pm$178 & 152.9 & 103.3 & 47.46\\ 
ZW1358.1+6245 & 0.328 & $0.315_{-0.003}^{+0.005}$ & $7.01_{-0.26}^{+0.25}$ & $0.43_{-0.07}^{+0.07}$ & 18322$\pm$149 & 123.2 & 88.6 & 51.1\\ 
MACSJ0404.6+1109 & 0.355 & $2.146_{-0.048}^{+0.052}$ & $44.51_{-8.76}^{+12.38}$ & $3.09_{-1.3}^{+2.46}$ & 2368$\pm$59 & 39.9 & 120.5 & 3.67\\ 
RXJ0027.6+2616 & 0.367 & $0.373_{-0.023}^{+0.024}$ & $8.08_{-1.40}^{+1.19}$ & $0.514_{-0.24}^{+0.29}$ & 702$\pm$31 & 22.7 & 81.2 & 5.15\\ 
MACSJ1720.2+3536 & 0.387 & $0.385_{-0.008}^{+0.005}$ & $6.51_{-0.36}^{+0.35}$ & $0.45_{-0.07}^{+0.08}$ & 6370$\pm$84 & 75.9 & 91.0 & 50.46\\ 
ZW0024.0+1652 & 0.395 & $0.397_{-0.005}^{+0.006}$ & $4.58_{-0.27}^{+0.5}$ & $0.76_{-0.17}^{+0.19}$ & 2671$\pm$59 & 45.3 & 59.0 & 29.0\\ 
RXJ1416.4+4446 & 0.400 & $0.413_{-0.016}^{+0.007}$ & $3.6_{-0.19}^{+0.22}$ & $0.87_{-0.21}^{+0.25}$ & 2120$\pm$51 & 41.9 & 76.3 & 31.91\\ 
MACSJ0159.8-0849 & 0.405 & $0.406_{-0.009}^{+0.01}$ & $9.54_{-0.5}^{+0.75}$ & $0.35_{-0.08}^{+0.08}$ & 7852$\pm$93 & 84.5 & 113.2 & 20.98\\ 
MACSJ2228.5+2036 & 0.412 & $0.411_{-0.009}^{+0.008}$ & $8.25_{-0.58}^{+0.59}$ & $0.35_{-0.08}^{+0.09}$ & 6004$\pm$86 & 69.4 & 137.8 & 18.47\\ 
MS0302.7+1658 & 0.424 & $0.427_{-0.014}^{+0.014}$ & $4.38_{-0.44}^{+0.6}$ & $0.45_{-0.19}^{+0.24}$ & 652$\pm$27 & 24.0 & 68.9 & 8.48\\ 
MACSJ0417.5-1154 & 0.44 & $0.458_{-0.011}^{+0.016}$ & $12.58_{-1.04}^{+1.3}$ & $0.33_{-0.11}^{+0.11}$ & 7544$\pm$95 & 79.4 & 132.8 & 9.84\\ 
MACSJ1206.2-0847 & 0.44 & $0.403_{-0.012}^{+0.014}$ & $11.79_{-0.71}^{+1.07}$ & $0.22_{-0.08}^{+0.08}$ & 11267$\pm$112 & 100.2 & 127.9 & 7.8\\ 
RXJ1701.3+6414 & 0.453 & $0.454_{-0.009}^{+0.008}$ & $4.49_{-0.25}^{+0.31}$ & $0.51_{-0.12}^{+0.13}$ & 2731$\pm$59 & 46.1 & 68.9 & 27.75\\ 
RXJ1641.8+4001 & 0.464 & $0.45_{-0.01}^{+0.008}$ & $4.64_{-0.43}^{+0.64}$ & $0.54_{-0.19}^{+0.22}$ & 983$\pm$35 & 28.4 & 46.7 & 12.91\\ 
MACSJ1824.3+4309 & 0.483 & $0.368_{-0.01}^{+0.013}$ & $6.2_{-0.96}^{+1.47}$ & $0.89_{-0.35}^{+0.48}$ & 469$\pm$26 & 17.9 & 78.7 & 7.9\\ 
MACSJ1311.0-0311 & 0.494 & $0.507_{-0.015}^{+0.02}$ & $8.79_{-0.94}^{+1.08}$ & $0.39_{-0.14}^{+0.15}$ & 2083$\pm$48 & 43.6 & 78.7 & 7.8\\ 
RXJ1524.6+0957 & 0.516 & $0.523_{-0.014}^{+0.013}$ & $5.61_{-0.48}^{+0.64}$ & $0.34_{-0.13}^{+0.14}$ & 1974$\pm$52 & 37.7 & 73.8 & 8.63\\ 4.6+0957
MS0015.9+1609 & 0.541 & $0.551_{-0.008}^{+0.009}$ & $8.37_{-0.3}^{+0.32}$ & $0.32_{-0.05}^{+0.05}$ & 16826$\pm$139 & 121.0 & 118.1 & 49.81\\ 
MACSJ1423.8+2404 & 0.543 & $0.557_{-0.013}^{+0.012}$ & $7.75_{-0.53}^{+0.71}$ & $0.31_{-0.09}^{+0.1}$ & 3518$\pm$62 & 57.2 & 76.3 & 14.07\\ 
MACSJ1149.5+2223  & 0.544 & $0.544_{-0.012}^{+0.012}$ & $12.75_{-0.96}^{+1.17}$ & $0.27_{-0.1}^{+0.1}$ & 9095$\pm$109 & 83.3 & 150.1 & 7.13\\ 
SC1120-1202 & 0.562 & $0.438_{-0.016}^{+0.014}$ & $5.36_{-0.74}^{+0.91}$ & $0.57_{-0.25}^{+0.32}$ & 714$\pm$33 & 21.7 & 51.7 & 6.48\\ 
MS2053.7-0449 & 0.583 & $0.595_{-0.019}^{+0.019}$ & $6.95_{-0.59}^{+0.67}$ & $0.2_{-0.11}^{+0.11}$ & 1884$\pm$49 & 38.5 & 49.2 & 3.69\\ 
RXJ0956.0+4107 & 0.587 & $0.02_{-0.012}^{+0.013}$ & $5.79_{-1.02}^{+1.56}$ & $0.79_{-0.45}^{+0.81}$ & 467$\pm$24 & 19.0 & 64.0 & 3.78\\ 
MACSJ2129.4-0741 & 0.589 & $0.617_{-0.015}^{+0.013}$ & $10.4_{-1.18}^{+1.37}$ & $0.57_{-0.14}^{+0.16}$ & 2918$\pm$59 & 49.1 & 100.9 & 18.33\\ 
MACSJ0647.7+7015 & 0.591 & $0.052_{-0.013}^{+0.013}$ & $7.49_{-0.61}^{+0.93}$ & $0.24_{-0.19}^{+0.16}$ & 2899$\pm$58 & 49.9 & 78.7 & 3.52\\ 
RXJ0542.8-4100 & 0.634 & $0.657_{-0.057}^{+0.035}$ & $8.09_{-0.95}^{+1.1}$ & $0.19_{-0.13}^{+0.14}$ & 1908$\pm$50 & 38.5 & 64.0 & 2.04\\ 
MACSJ0744.9+3927 & 0.698 & $0.728_{-0.013}^{+0.01}$ & $9.61_{-0.54}^{+0.77}$ & $0.34_{-0.08}^{+0.09}$ & 5669$\pm$79 & 71.4 & 81.2 & 17.68\\ 
RXJ1221.4+4918 & 0.70 & $0.629_{-0.014}^{+0.013}$ & $8.16_{-0.76}^{+0.84}$ & $0.35_{-0.11}^{+0.13}$ & 2828$\pm$61 & 46.6 & 73.8 & 10.39\\ 
RXJ2302.8+0844 & 0.722 & $0.64_{-0.026}^{+0.038}$ & $6.6_{-0.82}^{+0.92}$ & $0.26_{-0.15}^{+0.17}$ & 1387$\pm$44 & 31.8 & 46.7 & 3.27\\ 
RXJ1113.1-2615 & 0.725 & $0.773_{-0.015}^{+0.014}$ & $6.1_{-0.67}^{+0.83}$ & $0.56_{-0.18}^{+0.21}$ & 1135$\pm$38 & 29.9 & 39.4 & 13.42\\ 
MS1137.5+6624 & 0.782 & $0.855_{-0.03}^{+0.032}$ & $7.46_{-0.52}^{+0.59}$ & $0.34_{-0.11}^{+0.12}$ & 3957$\pm$67 & 59.1 & 44.3 & 10.59\\ 
RXJ1350.0+6007 & 0.804 & $0.834_{-0.019}^{+0.021}$ & $4.36_{-0.53}^{+0.8}$ & $0.65_{-0.27}^{+0.36}$ & 622$\pm$31 & 20.3 & 51.7 & 9.09\\ 
RXJ1317.4+2911 & 0.805 & $1.055_{-0.031}^{+0.035}$ & $3.84_{-0.68}^{+1.24}$ & $0.76_{-0.5}^{+0.8}$ & 230$\pm$20 & 11.4 & 29.5 & 3.05\\ 
RXJ1716.4+6708 & 0.813 & $0.834_{-0.009}^{+0.013}$ & $6.61_{-0.66}^{+0.75}$ & $0.61_{-0.16}^{+0.19}$ & 1341$\pm$40 & 33.8 & 44.3 & 19.28\\ 
MS1054.4-0321 & 0.832 & $0.834_{-0.02}^{+0.021}$ & $7.27_{-0.35}^{+0.48}$ & $0.23_{-0.07}^{+0.07}$ & 8637$\pm$103 & 83.8 & 68.9 & 12.46\\ 
WGA1226.9+3332 & 0.89 & $0.197_{-0.032}^{+0.027}$ & $7.25_{-0.69}^{+0.85}$ & $0.2_{-0.13}^{+0.15}$ & 2353$\pm$51 & 45.8 & 66.4 & 2.55\\ 
CLJ1415.1+3612 & 1.013 & $0.965_{-0.024}^{+0.034}$ & $6.31_{-0.6}^{+0.78}$ & $0.38_{-0.14}^{+0.16}$ & 1275$\pm$39 & 33.0 & 36.9 & 8.53\\ 
RXJ0910+5422  & 1.106 & $0.489_{-0.022}^{+0.026}$ & $4.1_{-0.56}^{+0.92}$ & $0.43_{-0.25}^{+0.36}$ & 419$\pm$24 & 17.2 & 24.6 & 3.69\\ 
RXJ1252-2927 & 1.235 & $1.226_{-0.03}^{+0.053}$ & $6.9_{-1.08}^{+1.2}$ & $0.84_{-0.28}^{+0.32}$ & 780$\pm$33 & 23.4 & 32.0 & 14.56\\ 
RXJ0849.9+4452 & 1.261 & $1.304_{-0.035}^{+0.054}$ & $4.3_{-0.71}^{+1.27}$ & $0.51_{-0.34}^{+0.48}$ & 327$\pm$22 & 14.6 & 22.1 & 2.87\\ 
XMM2235.3-2557 & 1.393 & $1.383_{-0.041}^{+0.046}$ & $9.6_{-1.12}^{+2.0}$ & $0.41_{-0.21}^{+0.24}$ & 1433$\pm$45 & 31.7 & 24.6 & 4.25\\ 
\hline
\end{tabular}
\caption{\label{fit}Best-fit results for our sample of Chandra
  clusters obtained with a free redshift parameter.  The focus is on
  the best-fit X--ray redshift $z_X$, compared to the optical value
  $z_o$.  We also list the X--ray temperature $T(z_X)$ and the
  abundance $Z_{Fe}(z_X)$.  Net photons (0.5-7 keV band) and
  S/N within the extraction radius $R_{ext}$ are
  also shown.  The value of $\Delta C_{stat}$ corresponding to the
  best-fit $z_X$ value is shown in the last column.}
\end{table*}

\subsection{Tuning the energy band}

The $K_\alpha$ Fe line is always located in the 2-7 keV energy range
(observing frame).  Therefore, before analyzing in detail our results,
we investigate whether focusing on this band may only help in driving 
the best-fit redshift value more efficiently towards the real one.  We can
either select the extraction region from the hard band image, or fit
the spectrum in the hard energy band only, or adopt both criteria.
The hard-band selected regions are usually smaller since the bulk of
the X-ray photons from the ICM thermal spectrum are in the soft (0.5-2
keV) band.  To compare the results of different strategies,
we introduce a quality parameter defined as $Q \equiv \sum
[(z_{X,i}-z_{o,i})/\sigma_{z_X}]^{2}/n$, where the sum is performed for the
entire sample (46 clusters) without removing the catastrophic
failures.  The value of $Q$ is a useful estimate of the average
discrepancy between $z_X$ and $z_o$ obtained with different
algorithms.  The comparison is shown in Table \ref{AB}, where we also
list the $\Delta z_{rms}$ values after applying $3\sigma$ clipping, and
the number of catastrophic failures.  We find that our default choice
based on the total energy band (0.5-7 keV) has a clear advantage with
respect to strategies focusing on the hard band.  We argue that the
signal in the soft band is useful for defining at best the continuum, and
this, in turn, positively affects the detection of the Fe line.  We therefore
conclude that there is no gain in restricting the energy band to the
2-7 keV range, and that the full amount of spectral information is
useful in searching for the Fe K$_\alpha$ line.

\begin{table}
\centering
 \begin{tabular}{c|c|c|c|c}
\hline
Region & Energy band & $Q$ & $\Delta z_{rms}$  & \# of cat. errors\\
\hline
Total & Total & 133.60 & 0.0288 & 8 \\
Total & Hard & 252.79 & 0.0378 & 7\\
Hard & Total & 284.14 & 0.1736 & 6\\
Hard & Hard & 468.66 & 0.0356 & 11 \\
\hline
\end{tabular}
\caption{Quality parameter for different choices of the energy band
  and typical rms value of $\Delta z$ after applying the $3\sigma$
  clipping.}
\label{AB}
\end{table}

\subsection{Dependence on the net detected photons}

\begin{figure}
\includegraphics[width=0.5\textwidth]{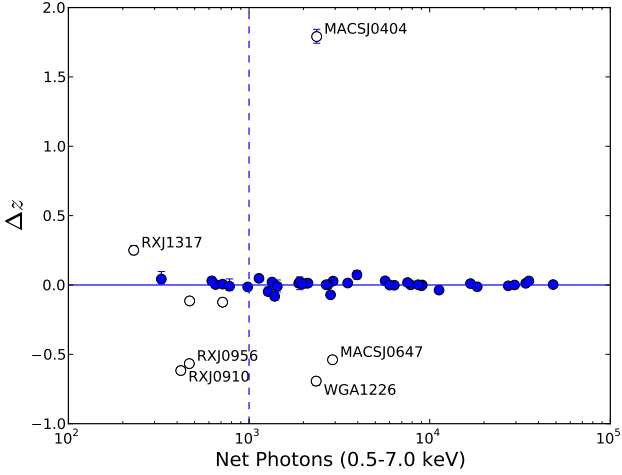} 
\caption{\label{cts}Difference between optical and X-ray redshifts as a
  function of the detected photons.  The eight empty circles are
  catastrophic failures. In some cases, spectra of high
  S/N (e.g. with more than 1000 net photons, shown with a vertical
  line) may not be successfully fitted. }
\end{figure}

We investigate how the average deviation and the number of
catastrophic failures depend on the S/N of the
spectra or, alternatively, on the total number of net detected
photons.  In Figure \ref{cts}, we show the quantity $z_X-z_o$ as a
function of the net detected photons.  From the inspection of Figure
\ref{cts}, we note that it is unclear whether a threshold on the number of
detected photons exists above which the X-ray redshift can be
considered reliable within a given confidence level.  When we consider 
S/N instead, a similar result is obtained.  We note that
above a threshold of about 1000 counts (vertical line) the number of
catastrophic failures is not negligible, although five out of eight
catastrophic failures are below this threshold.  Three catastrophic
errors occur in spectra with thousands of counts.  We draw the
conclusion that the efficiency of measuring the redshift depends not
only on the high quality of the signal, but also on the intrinsic
properties of the source. The likelihood of obtaining a reliable
spectral characterization, including the redshift, does not depend only on 
the net number of photons. 
To find a robust method to select reliable X-ray redshifts, we proceed
with a deeper investigation of the spectral parameters.  The most
obvious parameter to consider is the actual Fe abundance.

\subsection{Dependence on the intrinsic metal abundance}

\begin{figure}
\includegraphics[width=0.5\textwidth]{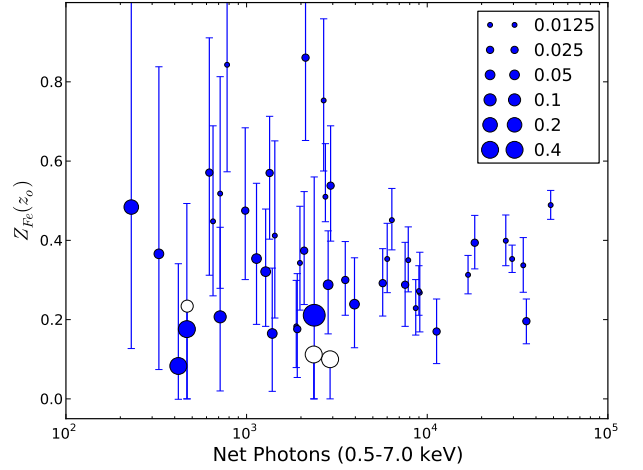} 
\caption{\label{ztn}''True'' Fe abundance (obtained by fitting the
  X-ray spectra freezing the redshift to the optical value) versus the
  total number of detected photons.  The radius of the dots is
  proportional to $|\Delta z|$. The three empty circles represent
    1 $\sigma$ upper limits.}
\end{figure}

We explore now whether the intrinsically low Fe abundance can be
another relevant source of error.  In Figure \ref{ztn}, we plot the
``true'' Fe abundance $Z_{Fe}$ (i.e., the value obtained when the
redshift is fixed to the optical value, see Table \ref{46}) versus the
number of net detected photons in each spectrum.  Most 
catastrophic failures are located in the bottom-left corner.  This
indicates that the largest discrepancies are associated with both the
low S/N of the X-ray spectrum and the low Fe abundance.
Significant discrepancies can be found among spectra with more than $\sim
4000$ photons if the Fe abundance is $Z_{Fe} \leq 0.2 Z_\odot$.

\begin{figure}
\includegraphics[width=0.5\textwidth]{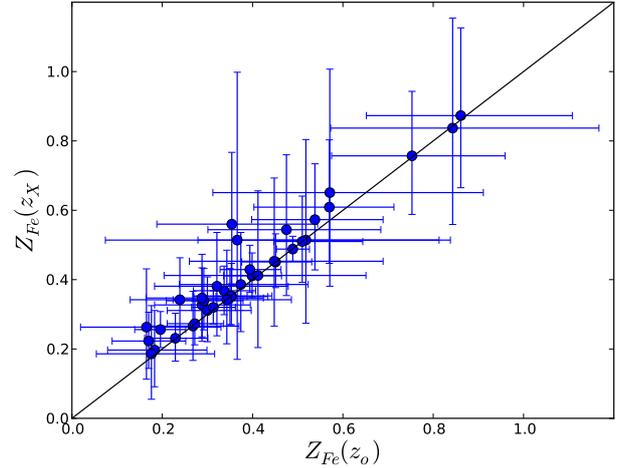} 
\caption{\label{zz}''True'' Fe abundances $Z_{Fe}(z_o)$ in the Chandra
  sample (obtained by fixing the redshift parameter to the optical
  value) plotted against those measured leaving the redshift parameter
  free. Catastrophic failures are not included.  The values of
  $Z_{Fe}(z_X)$ are systematically higher than those we derive when the
  redshift is fixed to the optical value.}
\end{figure}

\begin{figure}
\includegraphics[width=0.5\textwidth]{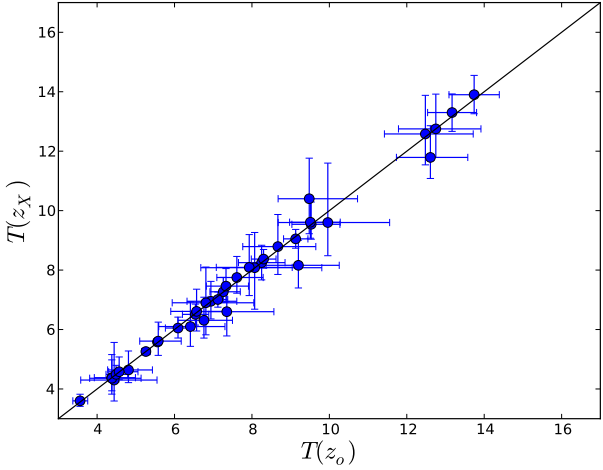} 
\caption{\label{TT}''True'' temperature $T(z_o)$ (obtained when the
  redshift parameter is fixed to the optical value $z_o$) plotted
  against the value $T(z_X)$ measured leaving the redshift parameter
  free.  Catastrophic failures are not included.}
\end{figure}

This information cannot be used in our blind search for the
emission lines, since we do not know a priori, the Fe abundance.  As
we see in the next section, making an assumption about the intrinsic
$Z_{Fe}$ value does not improve the fit.  In any case, we find that it
is impossible to use a sample whose redshift is determined via the
X-ray analysis to investigate the Fe abundance.  The X-ray fits with a
free redshift parameter will always lead to values that are
systematically higher than those obtained by fixing the redshift to
the optical value $z_o$, as shown in Figure \ref{zz}.  We find that
the typical value of the ratio of the fitted Fe abundance to the
"true" one is $\langle Z_{Fe}(z_X)/Z_{Fe}(z_o)\rangle = 1.110$, which
corresponds to a positive bias of 11\%. This result is expected since the
position of the Fe line is found by maximizing the Fe abundance for
a given temperature.  On the other hand, there is no evidence of a
bias in the best-fit values of the temperature, as shown in Figure
\ref{TT}.  For the temperature ratio, we find that $\langle
T(z_X)/T(z_o)\rangle = 0.993$.

\subsection{Dependence on $\Delta C_{stat}$}

\begin{figure}
\includegraphics[width=0.5\textwidth]{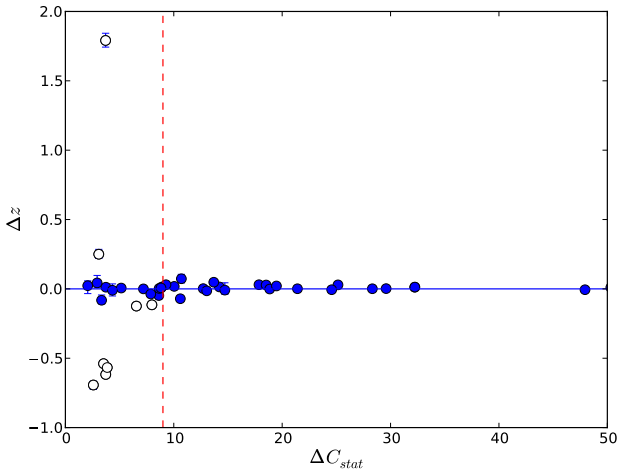} 
\caption{\label{dep}The quantity $\Delta z = z_X-z_o$ plotted against
  $\Delta C_{stat}$.  The eight empty circles are catastrophic failures.
  The vertical line show the nominal $3\sigma$ threshold corresponding
  to $\Delta C_{stat} = 9$.}
\end{figure}

Another important indication is the intensity of the emission line
itself.  The most reliable lines are those that provide the
largest decrease $\Delta C_{stat}$ in the C-statistics (see Figure
\ref{cstat}).  A larger $\Delta C_{stat}$ indicates a more robust
emission line, then a more reliable X-ray redshift, as shown in Figure
\ref{dep}.  Assuming that $\Delta C_{stat}$ behaves similarly to the
$\Delta \chi^{2}$ for one degree of freedom, we find that a threshold
that excludes all the catastrophic errors is $\Delta C_{stat} > 9$,
corresponding to a nominal confidence level of $3\sigma$.  Above this
level, we have 26 clusters for which the redshift is measured with
good accuracy, constituting a sample virtually free of catastrophic
errors.

\subsection{Exploring the secondary minima}

 We have investigated the properties of the absolute
  minima.  In principle, the $C_{stat}$ function of each cluster has
  several secondary minima that may contain useful information.  We
  explore this possibility by identifying all the local minima in the
  $C_{stat}-z$ function, and considering the corresponding $\Delta
  C_{stat}$ and $\Delta z$.  We find that, as for the catastrophic
  failure, none of their secondary minima provide a correct redshift 
  measurement.  As for the 38 clusters with acceptable redshift
  measurements, secondary minima do not provide any improvement in the
  redshift value, as expected.

We also explore the possibility of flagging unreliable $z_X$ from the
presence of one or more secondary minima close to the absolute
minimum.  However, applying a naive rejection based on 
more than one line of similar $\Delta C_{stat}$, would result in the
rejection of several successful cases.

In summary, we do not find any benefit when considering the
secondary minima, and we confirm that simply using a sharp threshold
on $\Delta C_{stat}$ is the most efficient criterion for selecting
reliable redshift measurements.

\section{Refining the strategy with weak priors}

We have previously here applied a direct blind search for the Fe line without
making any further assumption.  We now investigate whether we can
achieve a more efficient strategy using weak priors.  Two possible
choices of priors come from the study of local cluster samples: the low
scatter in the measured Fe abundance of hot clusters and the tightness
of the $L-T$ relation.

\subsection{Prior on the Fe abundance}

\begin{figure}[h]
\includegraphics[width=0.5\textwidth]{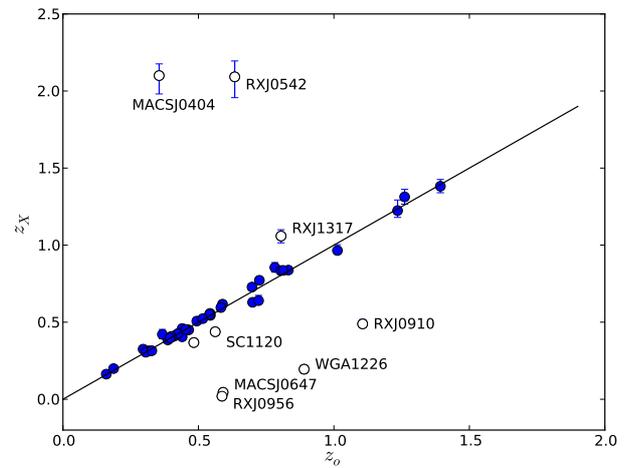} 
\caption{\label{ZOX}Same as Figure \ref{OX}, after assuming a fixed Fe
  abundance $Z_{Fe}=0.3 Z_\odot$.  }
\end{figure}

\begin{figure}[h]
\includegraphics[width=0.5\textwidth]{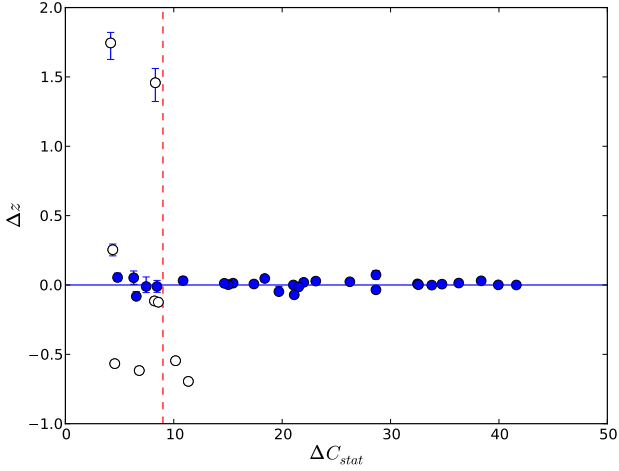} 
\caption{\label{zdep}Same as Figure \ref{dep}, after assuming a fixed Fe
  abundance $Z_{Fe}=0.3 Z_\odot$.  }
\end{figure}

The Fe abundance is observed to be almost constant in local and medium
redshift clusters for virial temperatures $kT > 5 $ keV
\citep{1997Renzini,2005Baumgartner,2007Balestra}.  This suggests that
we may remove one fitting parameter by freezing the Fe abundance to
$Z_{Fe} = 0.3 Z_\odot$.  This assumption would be wrong below 3 keV,
since in the low temperature range (from poor clusters to groups) the
Fe abundance values cover a wide range
\citep{1997Renzini,Rasmussen2007}.  To explore the effects of this
assumption, we repeat the fits with $Z=0.3 Z_\odot$ and show the
results in Figure \ref{ZOX}.  The $Q$ value after this assumption is
103.12, which represents a mild improvement in accuracy, mainly because of the
larger error bars of MACSJ0404 and RXJ0542. Incidentally, we note
that these two clusters with ``true'' Fe abundance below $0.2 Z_\odot$
are fitted with a redshift $\sim 2.1$, a mistake caused by an edge in
the Chandra response around 2 keV. The average rms value of $\Delta
z$ is also 0.03, and the number of catastrophic failures is 9.  We
conclude that there is no evidence that a prior on the Fe abundance
provide more reliable results. In addition, we find some indication that this
assumption might prevent us from being able to use the criterion about $\Delta
C_{stat}$ to select the most reliable redshift measurements, as
shown in Figure \ref{zdep}, where we have two catastrophic failures
for $\Delta C_{stat} > 9$.

\subsection{Priors on the $L-T$ relation}

Additional information is the clearly defined $L-T$ relation
measured for local groups and clusters of galaxies \citep[see, e.g.,
][]{2004XMM,2006Maughan,2009Pratt}.  Large errors in the redshift will
result in a position in the $L-T$ plane significantly distant from the
average observed relation, once the intrinsic scatter is properly
taken into account.

To illustrate this effect, we compute the bolometric luminosities for our
sample of clusters.  The luminosity contribution outside the extraction
regions is estimated by fitting a $\beta$ model to the observed
surface brightness.  All luminosities are computed for
$R_{500}$, which is defined as the radius within which the average
density contrast is 500 times the critical density of the universe 
at the cluster redshift. As for the L-T relation, we adopt the 
best fit by \citet{2006Maughan} to the Wide Angle ROSAT 
Pointed Survey (WARPS) sample at $0.6 < z < 1.0$:

\begin{equation}
 L_{500} = 4.97 \pm 0.80 \left( \frac{kT}{6keV} \right)^{2.80 \pm
   0.55} \: 10^{44} \, h_{70}^{-2} \, erg s^{-1} \, .
\end{equation} 

\begin{figure}
\includegraphics[width=0.5\textwidth]{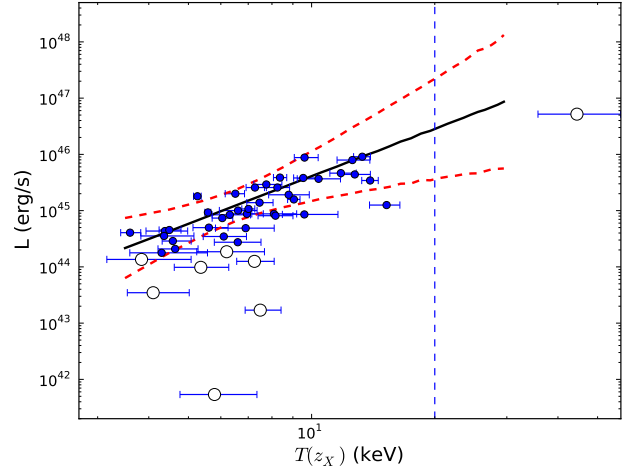}
\caption{\label{lt} $L$-$T$ relation for our cluster sample compared
  with the relation measured by \citet{2006Maughan}.  Catastrophic
  failures are marked as empty circles. The red dashed lines indicate
  the $3 \sigma $ confidence levels around the average $L$-$T$
  relation.  The vertical dashed line marks a reasonable upper limit
  (20 keV) that can be assumed for ICM temperatures.}
\end{figure}

In Figure \ref{lt}, we compare this relation with that obtained with
our cluster sample in a similar redshift range.  Catastrophic failures
are labeled with empty circles.  In total 14 clusters, including six
catastrophic errors, are formally excluded if we require all the
clusters to be consistent at the $3 \sigma$ level with the assumed
$L$-$T$ relation.  We note, indeed, that the catastrophic errors are
generally distant from the measured $L$-$T$ relation, but still very
close to the bulk of the cluster, so that a clear separation is not
observed.  This happens because the fitted temperature scales as
$(1+z_X)$, and the relation is approximately $L\propto T^3$.  This
implies that a wrong redshift would move the cluster luminosity
approximately along the observed $L-T$ relation.  Another less
relevant constraint that we show in Figure \ref{lt} is an upper limit on
the ICM temperature, which is conservatively put to 20 keV
(corresponding to the $3-8 \times 10^{15} M_{\odot}$ range).  This
constraint is expected to be useful in only a few extreme cases.

In principle, this prior should be applied at different redshifts because
the $L-T$ relation and its intrinsic scatter are expected to evolve at
some level.  However, there is still significant uncertainty in the
measured $L$-$T$ evolution \citep[see
][]{2004Ettori,2006Maughan,2007Branchesi}.  To summarize, the $L$-$T$
prior can be effective given that slope, normalization, and intrinsic
scatter are accurately known at different redshifts.  In practice,
this kind of information is expected to be obtained from cluster
samples as large as those achievable with future wide area surveys.
In this case, this criterion would not be a prior, but rather a
self-calibration procedure.  Therefore, with the present knowledge
this prior should be applied with caution.

\section{Building a complete sample of clusters with measured $z_X$: 
application to future X-ray surveys}

The work presented so far shows the efficiency of a blind search for
a redshift by means of the X-ray spectral analysis of known clusters
observed by Chandra.  On the basis of these results, we can define
general criteria for measuring accurate redshifts when applying this
procedure to a sample of new clusters, whose redshift is not known.
We can apply this to new extended sources detected in the Chandra and
XMM archives.  However, serendipitous searches in Chandra and XMM
pointings, which are inevitably confined to relatively small sky
areas, result in detections with a limited amount of net counts, such
as the ChaMP Galaxy Cluster Survey \citep{2006Champs}, the XMM-Newton
Distant Cluster Project \citep[XDCP, ][]{Fassbender}, the XMM-Newton
Large-Scale Structure \citep[XMMLSS, ][]{2007XMMLSS}, and the XMM
Clusters Survey \citep[XCS, ][]{XCS,2010Lloyd-Davies}.  In this case,
the application of our procedure will still be helpful but have a
modest impact.

The situation radically changes if we consider future X-ray surveys.
The planned eROSITA satellite \citep{2010Predehl,2010Cappelluti} and
the proposed WFXT mission \citep{2010WFXT} will provide a large number
of new detections.  We focus here on specific expectations for WFXT,
which is the most optimized mission for surveys thanks to its
wide-field optical design. This mission will be able to provide a
golden sample of 15000-20000 rich clusters, with $kT > 3 $ keV out to
$z \sim 1.5$, detected with more than 1500 counts \citep{2010Borgani}.

\subsection{Selecting clusters with reliable $z_X$}

We use the simulations we performed to assess WFXT science cases as
described in \citet{2010WFXTsimu}. Using the halo mass function by
\citet{Sheth1999}, we first determine the cosmological parameters
that most accurately fit the observed X-ray cluster luminosity function
\citep[XLF, see ][]{1998Rosati,2002Giacconi,2002Bauer} for a given
relation between cluster mass and X-ray luminosity. In this way, we
can extrapolate the observed XLF over the luminosity and redshift
ranges expected to be covered by the WFXT surveys. This luminosity
function is then used to generate mock realizations of the WFXT
cluster surveys through a Monte-Carlo procedure.  The corresponding
virial temperature (temperature gradients are not considered) is
assigned with a scatter of 10\% with respect to the mass, and the
luminosity with a scatter of 15\% - 45\% (going from massive clusters
to groups) with respect to the temperature, to be consistent
with the observed $L$-$T$ relation of local clusters.  We are thus
able to simulate an accurate spectrum normalized to the predicted flux
of each cluster.  Metal abundances are assigned randomly with a
Gaussian distribution centered around the average value $Z= 0.3
Z_\odot$ with a sigma $\Delta Z = 0.1 Z_\odot$, with a hard lower
limit at $Z = 0.1 Z_\odot$.  Values below $Z=0.1 Z_\odot$ have
never been measured in hot clusters at any redshift.  In addition,
stacked spectra analyses of distant clusters identify significant
enrichment in the ICM up to $z\sim 1.3$
\citep{2007Balestra,2008Maughan}.  As a matter of fact, a safe
assumption would be a constant Fe abundance equal to $0.3 Z_\odot$,
as also suggested by Figure \ref{ztn}.  However, we believe that by
allowing $Z_{Fe}$ values as low as $0.1 Z_\odot$, we will provide a
conservative estimate of the number of line detections.  We also
note that with this choice, in the simulation analysis we obtain
several best-fit values of the $Z_{Fe}$ parameter populating the
range below $0.1 Z_\odot$.  At present, it is impossible to provide
a more accurate model for the distribution of $Z_{Fe}$ in the ICM as
a function of redshift.

The background is computed accordingly to expectations for the
WFXT mission.  The net count rates per $deg^{2}$ of the different
background components are given in Table \ref{wfxtbck}.  We assume 
that the source spectrum is extracted from a 500 kpc region, a value
calibrated on the analysis of real Chandra data as in
\citet{2007Balestra}.  We also take into account that typically 10\%
of the total emitted flux is lost outside the extraction regions.  The
effect of vignetting is also considered, since the clusters will be
randomly positioned across the field of view.  This is an important
factor to be included because the vignetting is particularly severe in
the hard band, where the Fe line is located.  We use effective area
files corresponding to seven different off-axis angles in the range
0-30' to reproduce as closely as possible the vignetting effect
on the observed spectra.

\begin{table}
\centering
\begin{tabular}{ l|l l }
Background component & 0.5-2 keV & 2-7 keV\\
\hline
Particles & 0.188 & 0.397 \\
Galactic & 21.4 & 0.0 \\
AGN (Medium survey) & 3.9 & 1.65 \\
Cluster (Medium survey) & 0.79 &0.14 \\
\hline
Total (Medium survey) & 26.28 & 2.19 \\
\end{tabular}
\caption{\label{wfxtbck}WFXT background net count rates per $deg^{2}$
  in the Medium Survey (13.2 ks exposure).  The AGN and cluster
  contributions correspond to the emission of the unresolved AGN and
  groups/clusters, respectively, in the Medium Survey.}
\end{table}

We produce a mock catalog of groups (with temperatures above 0.5 keV)
and clusters as expected in 100 square degrees of the WFXT Medium
Survey \citep[see][]{2010WFXT2}.  The simulation thus consists of 100
pointings of 13.2 ks each, extracted from a total 3000 square degree
area.  We have about 2500 groups and clusters above the flux of $4
\times 10^{-15}$ erg s$^{-1}$ cm$^{2}$.  For all of them, we measure
the redshift $z_X$ with our procedure described in Section 3.  We remark
that this mock sample, being flux limited, includes a large number of
clusters with $kT \leq 2 $ keV, for which the presence of the L-shell
line complex at low energy significantly facilitates the measurement of
$z_X$.

\begin{figure}
\includegraphics[width=0.5\textwidth]{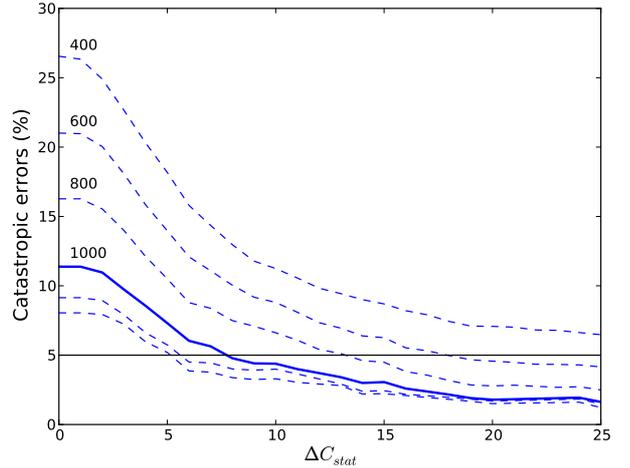}
\caption{\label{catastr}Percentage of catastrophic errors as a
  function of $\Delta C_{stat}$ in a sample with different
  net photons within $R_s$ from top to
  bottom.  The horizontal line marks the 5\% level.}
\end{figure}

\begin{figure}
\includegraphics[width=0.5\textwidth]{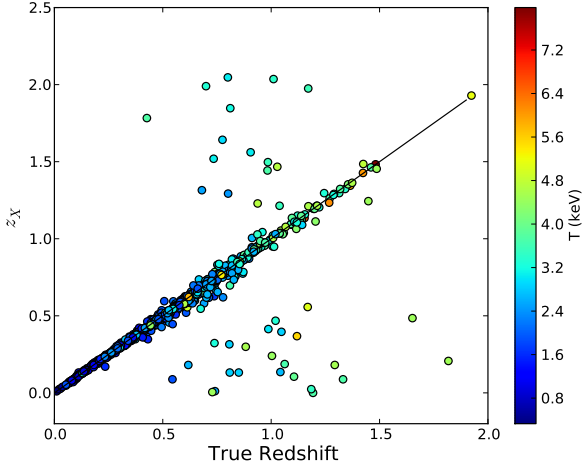}
\caption{\label{simOX1}X-ray measured redshift $z_X$ for all the
  clusters in a 100 square degree field of the WFXT Medium Survey
  detected with more than 1000 photons within the extraction region.
  We have 1037 clusters and groups, and 118 catastrophic failures.  The
  colors are set according to the cluster temperature.}
\end{figure}

\begin{figure}
\includegraphics[width=0.5\textwidth]{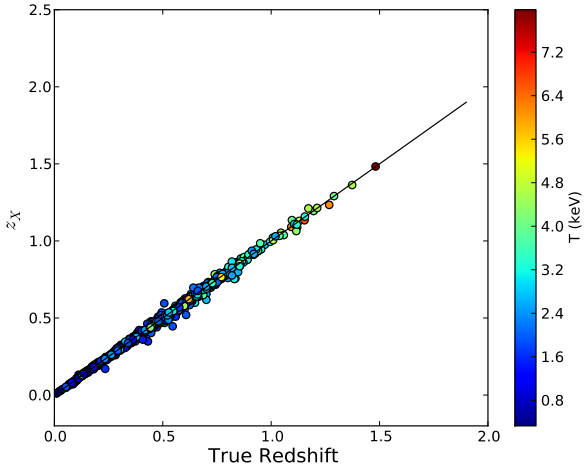}
\caption{\label{simOX2}The same as Figure \ref{simOX1}, but after
  applying the criterion $\Delta C_{stat} > 9$.  There are no
  deviations larger than 0.1, and only 38 clusters with $\Delta z >
  0.032$.  The size of the sample has decreased to 862.}
\end{figure}

\begin{figure}
\includegraphics[width=0.5\textwidth]{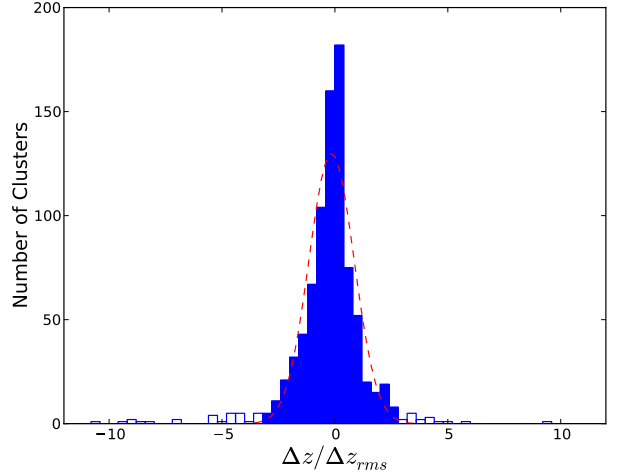} 
\caption{Histogram of $\Delta z/\Delta z_{rms}$ from a simulated 100
  square degrees of the WFXT Medium Survey. The sample here include all
clusters with more than 1000 photons and $\Delta C_{stat}$ larger than 9. 
Empty histogram indicates the residual catastrophic errors.
The red dashed line shows the Gaussian fit after $3\sigma$ clipping.}
\label{simhst}
\end{figure}

The catastrophic errors are identified by means of $3\sigma$ clipping.  In
Figure \ref{catastr}, we plot the percentage of catastrophic failures
as a function of $\Delta C_{stat}$ for clusters detected above a given
photon threshold.  We see that, for a given $\Delta C_{stat}$, the
number of catastrophic failures rapidly increases below 1000 net
photons.  By selecting a reference sample with more than 1000 net
photons and $\Delta C_{stat} > 9$, we are able to maintain the percentage of
catastrophic errors below 5\%.  The effect of the criterion $\Delta
C_{stat} > 9$ can be appreciated by comparing Figs \ref{simOX1} and
\ref{simOX2}.  In Fig. \ref{simOX1}, we show all the groups
and clusters with more than 1000 net photons, while in Fig.\ref{simOX2}
we apply the threshold $\Delta C_{stat} > 9$, which helps to define a
``golden'' sample of 862 groups and clusters.  This sample still
includes 38 catastrophic errors (4.4\% of the total) but all of them
are rejected marginally with redshift deviations slightly larger than
0.032. After applying the cut for $\Delta C_{stat} > 9$, the mean
redshift offset $\langle \Delta z \rangle $ is -0.0024 and the
deviation rms $\Delta z_{rms}$ is 0.0116 (see Figure \ref{simhst}).
We note that, despite 38 measurements being classified as catastrophic
errors, their typical discrepancy is still very small, and the overall
redshift accuracy of the sample with $\Delta C_{stat} > 9$ satisfies the 
demands of cosmological tests.  For example, in order not to degrade
dark energy constraints by more than 10\%, both $\langle \Delta z \rangle <
0.003$ and $\Delta z_{rms} < 0.03$ are required \citep{2007Lima}, which
are satisfied by our ``golden'' sample.

\begin{figure}
\includegraphics[width=0.5\textwidth]{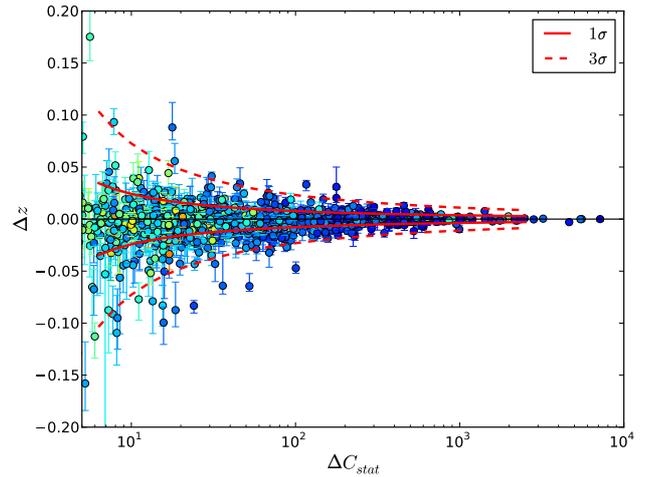} 
\caption{\label{simdep}$\Delta z = z_X-z_o$ versus $\Delta C_{stat}$
  for clusters with more than 1000 photons.  Error bars show $1\sigma$
  statistical error from XSPEC fitting.  The color code is the same as in
  Figure \ref{simOX1}.  There is an obvious trend of $\Delta z$
  decreasing at larger $\Delta C_{stat}$.  The red solid lines delineate
  $1\sigma$ envelope for $\Delta z_{rms}$. The red dashed lines
  delineate the $3\sigma$ confidence levels.}
\end{figure}

 Figure \ref{simdep} shows $\Delta z = z_X-z_o$ as a function of
  $\Delta C_{stat}$ for all the clusters with detections of more than 1000 photons.
The error rapidly decreases with increasing $\Delta C_{stat}$, as
expected.  The relation between $\Delta z_{rms}$ and $\Delta C_{stat}$
can be approximated as

\begin{equation}
\Delta z_{rms} = 0.025(log_{10} \Delta C_{stat})^{-1.5} \, .
\label{err_dc}
\end{equation}

\noindent
With this information we can select a sample with reliable redshifts and a
robust error estimate for $z_X$.  

\subsection{Assessing the completeness of the sample}

Samples selected with source counts above a given number of net
photons can still be treated as "complete".  Such a
threshold corresponds to a good accuracy for the limiting flux as a
function of the position on the sky, once the exposure time and the
vignetting in each field of view is computed.  Therefore, it is possible
to obtain a clearly defined sky coverage from which the XLF and the
X-ray Temperature Function (XTF) can be computed for cosmological
applications.

However, it is clear that a criterion based solely on the net counts
is an insufficient one.  As is immediately visible from Figure
\ref{simOX1}, among the clusters whose X-ray spectra have more than
1000 photons, there are about 120 catastrophic failures, several of
which are at a redshift much higher than the true one.  This implies 
that there may be a significant contamination at the bright ends of the
luminosity and temperature functions, which are both particularly sensitive
to cosmological parameters.  For example, six clusters in our sample
with more than 1500 counts are incorrectly located at $z>1.5$. For the full
$3000$ deg$^2$ Medium Survey, this would correspond to about 180 fake $z>1.5$
clusters, compared to the 20 expected \citep[see Figure 2
  in][]{2010Borgani}.  This implies that the majority of the rarest
$z>1.5$ clusters candidates would be spurious, and therefore that the
constraints on both cosmological parameters and non-Gaussianity
\citep[e.g.][and references therein]{2010Sartoris,2010Verde} would be
highly biased without applying the $\Delta C_{stat} > 9$ criterion.

\begin{figure}
\includegraphics[width=0.5\textwidth]{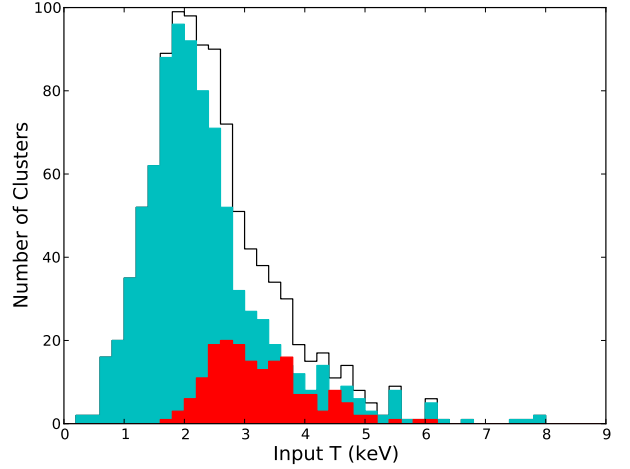}
\caption{Temperature distribution for the sample of groups and
  clusters detected with more than 1000 net photons and $\Delta
  C_{stat} > 9$ (filled histogram).  The black line indicates the
  complete input catalog, while the red histogram refers to the
  clusters removed from the flux-limited sample after applying the
  criterion $\Delta C_{stat} > 9$.}
\label{badhisto}
\end{figure}

By adopting the threshold on $\Delta C_{stat}$, the price to pay is a
lower level of completeness of the sample.  In the aforementioned
simulation, we have excluded 175 clusters (17\% of the total) detected
with a number of photons above the threshold because they do not
satisfy the $\Delta C_{stat} > 9$ criterion.  In this way, we remove
most of the catastrophic errors but also lose about 100
clusters with a good $z_X$.  As a result, the "golden sample" is 
incomplete above a given flux threshold, and the missing clusters must
be accounted before applying cosmological tests.
Unfortunately, this step is difficult because the temperature
distribution of the clusters with $\Delta C_{stat} <9$ is biased with
respect to the distribution of the entire sample.  This is shown in
Figure \ref{badhisto}, where the normalized distribution of rejected
clusters with more than 1000 photons is skewed towards high
temperatures.  This occurs because it is generally more difficult to
fit the Fe line in increasingly hot clusters, since the abundance of
He-like Fe ions, which dominate the line emission, begins to decrease
at temperatures $kT > 5$ keV, and the thermal continuum steadily
increases.  Such a bias against the most massive objects needs to be
carefully quantified. 

There are two ways to restore the completeness of the sample.  One is
to statistically correct the incompleteness using results from
N--body simulations.  However, this is affected by serious systematic errors
because we modify the high-mass end of the cluster distribution,
which is the most sensitive to cosmological parameters.  The other
option is to limit spectroscopic follow-up to only those clusters
with $\Delta C_{stat} < 9$.  

To design the optimal strategy, we should proceed with a
detailed investigation of the effects of varying the threshold of
$\Delta C_{stat}$ on the cosmological parameter constraints.  We
noted in Section 5.1 that $\Delta C_{stat} >9$ provide a "golden
sample" that largely satisfies the requirements of achieving a
precision of 10\% for the dark energy constraints.  Relaxing the
threshold on $\Delta C_{stat}$ and mapping the threshold values to
the accuracy on the cosmological parameter is clearly an option that
we plan to explore in a future paper.

Another possible source of contamination may be diffuse X-ray 
emissions produced by inverse Compton processes associated with radio
jets \citep[see, for example,][]{2003Fabian}.  These sources are a
serious contaminant when detecting X-ray extended sources,
especially at very low fluxes. However, this source will hardly
provide a spurious line detection above the selection threshold, 
hence their presence in a cosmological "golden" sample can be
neglected. 

\section{Conclusions}

We have performed a blind search of the Fe line in the X-ray spectra of
clusters of galaxies using Chandra archival data.  Our goal is to
define the optimal methodology for determining X-ray redshifts and to quantify
their accuracy by investigating both statistical and systematic errors.
To this end, we compared $z_X$ with the value $z_o$ determined with
optical spectroscopic observations.  

We have found good agreement in general, but also in several cases where
  $z_X$ is determined by the false detection of an emission line.
Thus we explored different methods to minimize the number of these
catastrophic failures.  We have found that it is preferable to use the total available band
(0.5-7 keV) for both selecting the extraction region and fitting
the spectra.  Catastrophic failures are found to be
caused by low spectrum S/N and an intrinsically low Fe
abundance.  For this reason, a lower limit to the net detected
  photons is insufficient to guarantee a robust measurement of the
  redshift.  However, we find that we can exclude most 
catastrophic errors by using the criterion $\Delta C_{stat}> 9$, where
$\Delta C_{stat}$ is the difference between the best-fit value of
$C_{stat}$, and its best-fit value when the metal abundance is
constrained to be zero.  

We have also explored the use of weak priors to improve the
results.  We have found that by fixing the Fe abundance to the local average
value $Z_{Fe} = 0.3 Z_\odot$, the number of catastrophic errors is not
reduced.  Furthermore, by requiring that all the fitted clusters 
agree with the observed local $L$-$T$ relation and its scatter,
one has an independent method to identify outliers, although not as
efficiently as the condition $\Delta C_{stat}>9$.

Our spectral analysis of Chandra clusters shows that future X-ray
survey missions will be able to define sizable samples of clusters
with X-ray measured redshifts. We specifically investigated the case
of the Wide Field X-ray Telescope using a mock catalog of $\sim 2500$
groups and clusters extracted from a 100 square degree area of the WFXT
Medium Survey.  The input cluster catalog is generated from the
cluster mass function normalized to the observed space density of
X--ray clusters by assuming local scaling relations to derive X--ray
luminosities and temperatures associated with a given mass. We found
that, by applying the condition $\Delta C_{stat}>9$, we can
successfully measure the redshift of 862 clusters with more than
1000 net counts, which leaves a small fraction (4.4\%) of marginally
catastrophic errors.  We  argue that these subsamples with
X--ray determined redshifts can be effectively used for cosmological
applications, thus avoiding time-consuming spectroscopic observations,
although additional simulations will have to be developed to assess the completeness of the
sample down to a given flux limit.

\acknowledgements

We thank Stefano Ettori for helpful discussions, Sergio
Campana, Paolo Conconi, and Andy Ptak for providing the WFXT response
files.  We also thank the anonymous referee for insightful comments and
suggestions.  We acknowledge financial contribution from contract
ASI--INAF I/088/06/0, from the PD51 INFN grant, and from the PRIN-MIUR
grant "The cosmic cycle of baryons".  The work is also supported by
the National Science Foundation of China under the Distinguished Young
Scholar Grant 10825313， the Ministry of Science and Technology
national basic science Program (Project 973) under grant
No. 2007CB815401, and the Fundamental Research Funds for the Central
Universities.  HY acknowledges the support of ICTP-IAEA Sandwich
Training Educational Programme (STEP), and of the China Scholarship
Council.

\bibliography{ref}

\end{document}